# A Novel Service Deployment Policy in Fog Computing Considering The Degree of Availability and Fog Landscape Utilization Using Multiobjective Evolutionary Algorithms

Maryam Eslami
*Dept. of Computer Engineering*
*Bu-Ali Sina University*
Hamedan, Iran
m.eslami@eng.basu.ac.ir

Mehdi Sakhaei-nia
*Dept. of Computer Engineering*
*Bu-Ali Sina University*
Hamedan, Iran
sakhaei@basu.ac.ir

*Abstract—* Fog computing is a promising paradigm for real-time and mission-critical Internet of Things (IoT) applications. Regarding the high distribution, heterogeneity, and limitation of fog resources, applications should be placed in a distributed manner to fully utilize these resources. In this paper, we propose a linear formulation for assuring the different availability requirements of application services while maximizing the utilization of fog resources. We also compare three multiobjective evolutionary algorithms, namely MOPSO, NSGA-II, and MOEA/D for a trade-off between the mentioned optimization goals. The evaluation results in the iFogSim simulator demonstrate the efficiency of all three algorithms and a generally better behavior of MOPSO algorithm in terms of obtained objective values, application deadline satisfaction, and execution time.

## I. INTRODUCTION

The fact that most IoT devices have hardware and energy limitations, enabling computation technologies is required for processing relevant data. Over the past decades, cloud computing has been one of the facilitating technologies for different IoT scenarios [1]. However, most new applications such as smart surveillance, smart transportation, and smart healthcare, have specific requirements such as real-time/near-time processing, location awareness, etc. Despite many significant advantages, cloud computing as a more centralized model is not a proper choice for such scenarios. In this regard, various research efforts have been founded to propose more distributed models in order to overcome the problems with the central nature of cloud computing [2].

Fog Computing was introduced by CISCO in 2012 with the promises of latency reduction, location awareness, support for mobility, and support for heterogeneity [3]. The vision of fog computing, as the name implies, is extending cloud services hierarchically to the edge of the network aiming at reducing the distance between data sources (IoT devices) and data centers.

Fog computing acts as the processing core of different IoT applications. In cloud-based scenarios, each application is deployed on a VM instance that runs on a physical server. This model, however, is not suitable for fog. Fog resources are much more limited in terms of hardware capabilities so deploying an application with all its functionalities and then managing the entire workload is not possible in this arena. This, however, does not seem to be a drawback, as over the past years the application layout has shifted from monolithic to modular/micro-services, which fits exactly with a distributed computation model such as the fog. Therefore, each fog node hosts only those services/modules of applications that their required resources do not exceed its capacity. That's why deploying applications in the fog is called *service deployment* [4], [5].

Typically, Fog Service Deployment (FSD) is modeled as an optimization problem with decision variables, constraints, and objective functions. Several optimization metrics have been considered in the literature for Fog Service Deployment Problem (FSDP) optimization. Various methodologies have also been used for finding the optimal solution. However, there is a lack of methodologies that take into account non-functional requirements of applications such as reliability, availability, security, etc. Although, in some applications due to mission-critical and safety-critical nature, these types of requirements need special attention and cannot be ignored easily.

Ensuring the availability requirements of application services while maximizing utilization of fog resources to meet their deadline requirements is a challenging task that has not gained significant attention in previous studies. To ad-dress this, we formulate the FSDP as an Integer Linear Programming (ILP) problem and compare the efficiency of three reputed multiobjective evolutionary algorithms namely, MultiObjective Particle Swarm Optimization (MOPSO) [6], Non-dominated Sorting Genetic Algorithm-II (NSGA-II) [7], and MultiObjective Evolutionary Algorithm based on Decomposition (MOEA/D) [8] for holding a trade-off between the two conflicting goals. To the best of our knowledge, this is the first study that considers the abovementioned metrics for service deployment optimization in the fog. On the other hand, adopting the multiobjective evolutionary algorithms is another innovative aspect of this paper.

The remaining sections of this paper are organized as follows: In Section II some of the most related previous studies are reviewed. The details of the system model and the proposed problem formulation are described in Section III. Experimental setups and evaluation results are stated in Section IV. Finally, the conclusion and some directives for future work are presented in Section V.

## II. RELATED WORK

Service deployment in fog computing has been considered extensively by the research community in most recent years. A summary of more related studies along with the considered metrics and optimization methodologies is given in Table I. It must be mentioned that for the sake of brevity only studies using evolutionary algorithms are included. But the fact is that in most researches heuristics of own are proposed for solving the FSDP. Regarding the optimization metrics, it is evident that time-related indicators such as service latency, network latency, response time, makespan, etc. have received attention in most of the researches. Energy and cost fill the second and the third place respectively. As can be seen, genetic algorithm has been used repeatedly in the literature as the optimization strategy. PSO algorithm is the second most used bio-inspired algorithm. The work of Natesha et al. in 2018 [9] maybe is the first attempt for multiobjective optimization of FSDP. Other multiobjecive evolutionary algorithms such as the NSGA-II, MOPSO, SPEA-II, MOEA/D, etc. are rarely used in the studies of the field.

As a mapping problem, the FSDP is supposed to be NP-Hard [10], i.e., finding the optimal solution is not possible in polynomial time with the adoption of exact mathematical methods. A quick look at the experimental results of the papers represents the high computational complexity of these methods, especially for the large-sized configurations as in the fog [11]–[14]. On the other hand, the efficiency of evolutionary algorithms to find near-optimal solutions in a timely manner has been proven in various optimization problems of different domains [15]. Nevertheless, these methodologies have not been yet adopted significantly in the fog service deployment studies. Another fact is that quality-related requirements of applications such as reliability, availability, security and privacy have been neglected unequally in the researches of the field. The work of Lera et al. [14] is maybe the only one that takes into account the availability along with the deadline satisfaction rate of applications. Their proposed method is based on the complex network theory that attempts to deploy interrelated services adjacent to each other by partitioning fog nodes and application services into communities and transitive closures respectively. They used the first-fit decreasing greedy algorithm for mapping the applications to the communities which is not an efficient choice for large-scale and complicated scenarios as in the fog. On the other hand, in their formulations, they did not investigate different availability requirements of application services and their impact on the suitability of a deployment strategy.

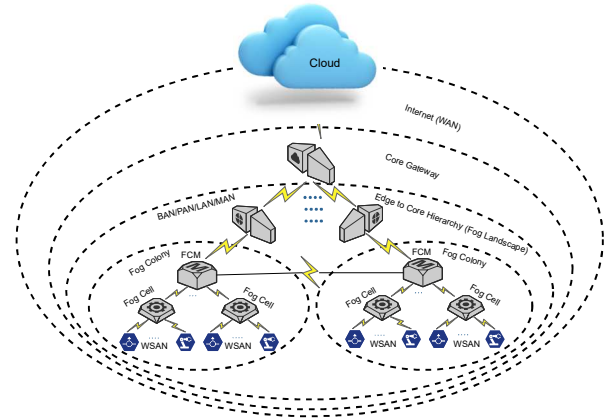

Fig. 1: Hierarchical Fog Computing Architecture

The details of the proposed ILP formulation to address the aforementioned research gaps are discussed consequently.

## III. SYSTEM MODEL AND PROBLEM FORMULATION

### A. Fog Architecture

The conceptual fog framework considered in this paper is based on the one proposed in [16]. This model addresses almost all main aspects of fog computing such as limitation, heterogeneity, ubiquitousness, location-awareness, and the interoperability with the cloud. Based on the proximity of the resources, the whole fog landscape is divided into fog colonies. In each colony, a node that poses more capabilities, acts as the Fog Colony Manager (FCM), i.e. regularly monitors the resources of the underneath Fog Cells (FCs), receives service requests, makes decision about locus of processing request, i.e. deploying services, communicates with the counterparts in neighbor colonies and the central cloud. Fig. 1 illustrates the general view of this framework.

In the abovementioned research, neighbor colonies are chosen based on the latency between the corresponding FCMs. To ensure the availability requirements of services, we extend this model by considering the failure history of the resources when making decisions about where to offload service requests.

### B. Application Model

To distribute application services over fog resources, programming model is considered as the Distributed Data Flow (DDF) that models an application as a Directed Acyclic Graph (DAG). Nodes of this graph are the application services, and the edges represent the data flow between them [25].

TABLE I: Summary of Fog Service Deployment Studies

| Reference | Considered Metric(s) | Methodologies |
|---|---|---|
| [16] | Fog Resource Utilization, Deadline of Applications | GA[a] |
| [14] | Service Availability, Deadline of Applications | First-Fit Decreasing Greedy Algorithm |
| [9] | Service Time, Energy, and Cost | Own method (based on the elitist GA) |
| [17] | Makespan, Energy | GA-PSO[b] |
| [18] | Makespan, Energy | Own method (based on the PSO) |
| [19] | Latency, Cost | Own method (based on the ACO[c]) |
| [20] | Response Time, Throughput, and Cost | Own method (based on the GWO[d]) |
| [21] | Response Time, Energy | Own method (based on the Memetic Algorithm) |
| [22] | Energy | Own method (based on PSO) |
| [23] | Response Time, Cost, and Energy | NSGA-II |
| [15] | Network Latency, Service Spread, and Resource Utilization | NSGA-II, MOEA/D, and WSGA[e] |
| [24] | Latency, Fog Resource Utilization, and Cost | SPEA-II[f] |
| This Work | Availability of Services, Deadline of Applications, and Fog Landscape Utilization | MOPSO, NSGA-II, and MOEA/D |

[a]Genetic Algorithm, [b]Combination of GA and Particle Swarm Optimization Algorithm, [c]Ant Colony Optimization, [d]Gray Wolf Optimization [e]Weighted Sum GA, [f] Strength Pareto Evolutionary Algorithm-II

## C. Fog Service Deployment Problem

Now it is time to formally define Fog Service Deployment Problem (FSDP). Considering the system model described above, a fog landscape is a set of IoT applications $Apps = \{App_1, App_2, ..., App_m\}$, each $App_i$ consisting of $k_i$ modules/services $S_i = \{s_1, s_2, ..., s_{k_i}\}$. On the other hand there exists a set of resources $FR = fr_1, fr_2, ..., fr_n, CR$ including the resources in the colonies $fr_i$ and the cloud $CR$. In this way, the FSDP is to find a mapping from $\bigcup_{i=\{1,2,...,m\}} S_i$ to $FR$.

## D. Optimizatio Model

We formulate FSDP as a bi-objective optimization problem with the goals of maximizing the utilization of fog resources as well as the availability of services in the landscape. It should be noted that these are two conflicting objectives, i.e., maximizing fog utilization means deploying services on the fog colonies as much as possible. On the other hand, cloud resources have higher degrees of availability due to complex failure recovery mechanisms equipped with. In other words, deploying more services on the fog to meet the deadline requirements of applications might cost lessening total availability. In this regard, a proper trade-off should be established between these two goals. In the following sections, the Integer Linear Programming (ILP) model of the optimization problem including decision variables, objective functions, and constraints is proposed.

### 1) Objevtice Functions:

*a) Fog Utilizatoin:* Fog Utilization: In the case of receiving an application request, four scenarios might occur. In the first case, the request submitted to the FCM could be processed on its own. Another case is that based on the system state, the FCM decides to distribute relevant services on the fog cells in the colony. When the entire colony is not capable of handling the request, it is forwarded to the FCM of the neighbor colony which is the third case. Lastly, even if the neighboring colony does not respond, the entire request is sent to the cloud. The binary decision variables $x_{fcm}, x_{fc}, x_{nfc}$, and $x_c$ represent each of which respectively. We consider fog utilization as the fraction of all services allocated to either FCM, FC, or NFC. Therefore, the first objective function is formulated in the (1).

$$\frac{1}{N}\sum_{i=1}^{m}\sum_{j=1}^{k_i} x_{fc}^{s_{i,j}} + x_{fcm}^{s_{i,j}} + x_{nfc}^{s_{i,j}} \quad (1)$$

where

$$N = \sum_{i=1}^{m} |S_i| \quad (2)$$

*b) Availability:* Availability of an application expresses the responsiveness of its services when needed. Depending on what functionalities a service offers, different availability degrees are required. For example, the failure of a component that preprocesses data before sending it to the cloud for permanent storage is much more tolerable than the case of an emergency patient alert service.

Assuming that the applications are well designed and have no inherent implementation flaws, their responsiveness is directly affected by the failure probability of the resources in which they are deployed. To numerally evaluate the availability suitability of a deployment strategy, a score is assigned to each service mapping by comparing the availability requirement of the service and the failure probability of the resources in the landscape. The availability of the cloud resources is set to a large enoght value (99.999%). Equations (3)–(5) show the mathematical representation of what was described.

$$\sum_{i=1}^{m} \frac{1}{|S_i|} \sum_{j=1}^{k_i} Availability\ Score\ (s_{i,j}, R_{s_{i,j}}) \quad (3)$$

$$Availability\ Score\ (s, R_s) = \begin{cases} 1 & Availability\ Req.(s) \leq \Pr_{UP}^{R} \\ 0 & O.W. \end{cases} \quad (4)$$

$$Pr_{UP}^{R} = 1 - Pr_{Failure}^{R} \quad (5)$$

where $R_{s_{i,j}}$ is the resource in which the service instance $s_{i,j}$ ($j^{th}$ service of $App_i$) has been deployed on it. The coefficient $\frac{1}{|S_i|}$ in (3) is for normalizing scores proportional to the number of services of each application, e.g., the value 2 of availability score for an application with 4 modules must not be equal to that of 10 modules.

### 2) Constraints:
We consider three types of constraints in the optimization model. The first, as given in (6) states that each service instance must be deployed on either one of the FC, FCM,

TABLE II: Simulation Settings

(a) Applications Deadlines

| Application | Deadline (s) |
|---|---|
| $App_1$ | 300 |
| $App_2$ | 60 |
| $App_3$ | 180 |
| $App_4$ | 240 |
| $App_5$ | 120 |

(b) Hardware and Availability Requirements of Services

| Type of Service | CPU | RAM | Size | Availability Ratio (%) |
|---|---|---|---|---|
| Sense | 50 | 30 | 10 | Random (80%–95%) |
| Process1 | 200 | 10 | 30 | Random (70%–95%) |
| Process2 | 200 | 20 | 30 | Random (70%–90%) |
| Process3 | 100 | 30 | 30 | Random (90%–95%) |
| Actuate | 50 | 20 | 10 | Random (95%–100%) |

(c) Resources Characteristics

| Type of Resource | CPU | RAM | Failure Probability (%) |
|---|---|---|---|
| Cloud | 200000 | 200000 | 0.001% |
| FCM | 1000 | 512 | 10% |
| FC | 250 | 256 | 20% |

NFC, or the cloud. The second one expresses what was mentioned earlier, i.e., the hardware requirements (CPU, RAM, and storage) of the services should not go beyond the capacity of the resources hosting them. This is what showed up in (7)–(9). To avoid degrading the performance caused by the resources overbooking, a percentage c of each resource type is preserved for system-related tasks. As the third, we take into account response times of applications that should be passed before the specified deadline of each (10).

$$\sum_{i=1}^{m}\sum_{j=1}^{k_j} x_{fc}^{s_{i,j}} + x_{fcm}^{s_{i,j}} + x_{nfc}^{s_{i,j}} + x_{c}^{s_{i,j}} = 1 \quad (6)$$

$$\sum_{i=1}^{m}\sum_{j=1}^{k_j} X_{CPU}^{s_{i,j}} \leq (1-c) \times CPU_{R_{s_{i,j}}} \quad (7)$$

$$\sum_{i=1}^{m}\sum_{j=1}^{k_j} X_{RAM}^{s_{i,j}} \leq (1-c) \times RAM_{R_{s_{i,j}}} \quad (8)$$

$$\sum_{i=1}^{m}\sum_{j=1}^{k_j} X_{Storage}^{s_{i,j}} \leq (1-c) \times Storage_{R_{s_{i,j}}} \quad (9)$$

$$\forall App_i \in Apps: RT_{App_i} \leq Deadline_{App_i} \quad (10)$$

Finally for the binary decision variables we have (11).

$$\forall App_i \in Apps, \forall s_j \in S_i: X_{s_j} \in \{0,1\} \quad (11)$$

In summary, the formulated FSDP becomes as follows.

$$Maximize\ (1), (3) \quad (12)$$

$$Subject\ to: (6) - (11)$$

### IV. EVALUATION AND DISCUSSION

The three evolutionary algorithms and the problem formulation have been implemented and examined in the iFogSim simulator [26]. The *Sense-Process-Actuate* is the main workflow considered for applications in the iFogSim. The simulation parameters for a fog landscape consisting of five applications, each with five services and a four-layer physical topology, are summarized in Table II.

We consider three determining factors to compare the algorithms, first the values of optimization objectives, for the different values of maximum iterations (evaluations). As the second, we consider the application deadline satisfaction. Last but not least, the execution time of algorithms for the different number of services has been investigated as the third evaluation factor. The experimental results for each case are illustrated in Figure 2, 3, and 4 respectively.

In the case of objective values, all three algorithms achieve at least 70\% of fog landscape utilization. The MOPSO and NSGA-II compete closely with each other especially for the number of evaluations above 600. The same pattern can be seen for the total availability. In general, MOPSO and NSGA-II hold a better trade-off between the objectives. It can be seen that in some points such as 1000 number of evaluations, the NSGA-II has the highest value for fog utilization while having the smallest for the total availability. The other similar cases are due to the conflicting nature of objectives, i.e., the greatest value for one is at the cost of degrading the other. To conclude, the MOPSO algorithm performs better in keeping values at the moderation.

The results of deadline satisfaction are very interesting. As mentioned, the time requirement of applications is included as a constraint in the problem formulation (10). However, the results of all three algorithms are as if it is intended as a minimization goal. The horizontal lines in Figure 3a represent the application deadlines while the tiny columns at the floor represent the response time of applications. In addition, for more clarity, the results of the response times are given in a separate diagram (Figure 3b). As can be seen, there is an anomalous value for the case of $App_3$. This is because the values assigned for the availability ratio of its services are smaller than the others that led to deploying most of its services on the fog cells (highest failure probabilities). As the considered service model for these types of resources is based on the M/D/1 queuing model, the average execution time gets higher than the other cases. To summarize, the MOPSO algorithm has resulted in better values compared to the other two methods.

Lastly, the execution time plots (Figure 4) show a linear growth with the increase of the number of application services which is as expected to be according to the time complexity of each algorithm. The order of execution times is also according

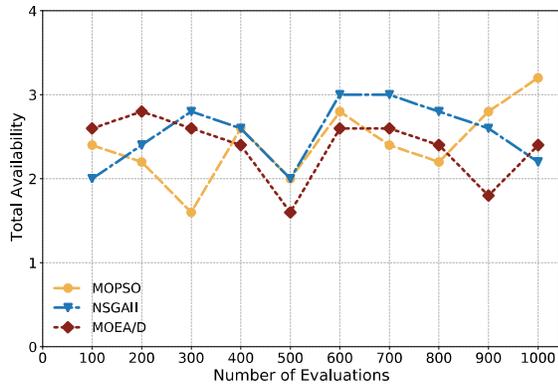 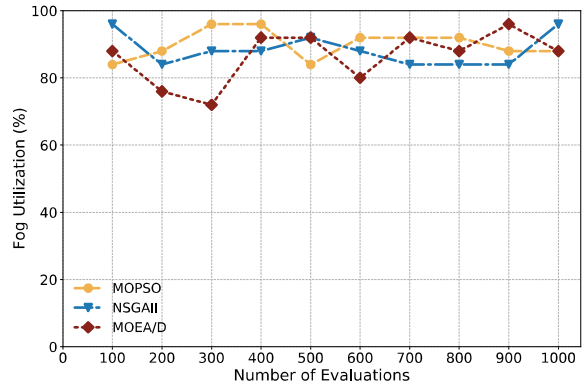

Fig. 2: Evolution Process of Algorithms

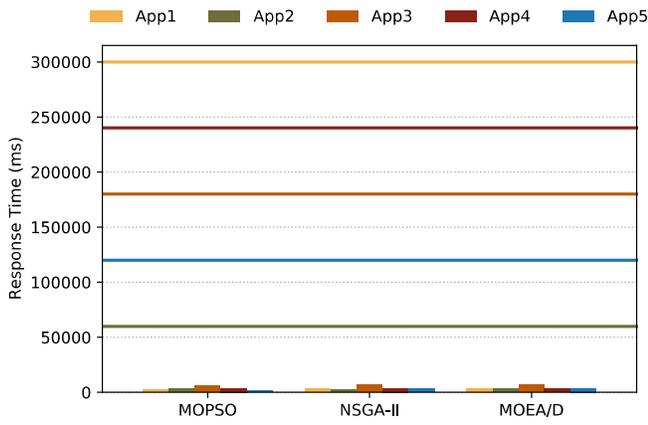 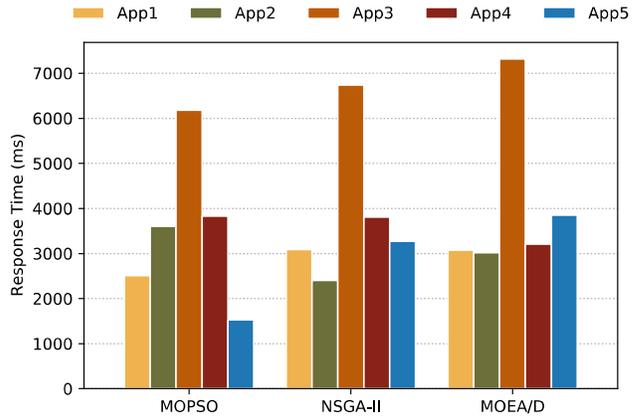

Fig. 3: Response Times and Deadlines of Applications

to the time complexity of each algorithm, i.e., NSGA-II and MOPSO have the highest and smallest execution times, respectively. The MOEA/D algorithm, on the other hand, takes a

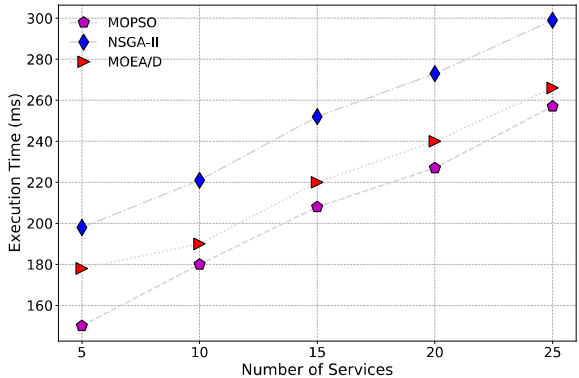

Fig. 4: Execution Times of Algorithms

## V. CONCLUSIONS AND FUTURE WORKS

Fog computing is maybe the next generation of computation paradigms that extends cloud services hierarchically to the edge of the network. Real-time and mission-critical Internet of Things applications are the most benefiting from the advantages of this computation model. Applications in fog-assisted scenarios are decomposed to lightweight services that can be hosted by constrained fog resources. To best deployment of services, both functional and non-functional requirements of applications must be taken into account. Despite the functional requirements such as time, cost, and energy non-functional requirements of applications have not been considered appropriately in the previous studies. To address this, in this paper we proposed a linear formulation considering the different availability requirements of services while maximizing utilization of fog landscape resources. MOPSO, NSGA-II, and MOEA/D algorithms were chosen for a trade-off between these two competing goals. The experimental tests in the iFogSim simulator resulted in high quality and near-optimal solution values for all three algorithms. Not in all cases, but generally, MOPSO performed better than the others regarding objectives values, deadline satisfaction, and execution time. Considering other non-functional requirements of applications such as security, along with the other metrics such as energy and cost might be a topic for future work. Also, adopting other multiobjective evolutionary algorithms such as SPEA-II, PESA-II, and bio-inspired swarm intelligence algorithms is another suggestion to continue this research.